\documentclass[11pt]{article}
\usepackage[utf8]{inputenc}
\usepackage{geometry}
\usepackage{footnote}
\usepackage{setspace}
\usepackage[english]{babel}
\usepackage{amsmath,amsthm,amssymb,xcolor}
\usepackage{smartdiagram}  
\usepackage{placeins}
\usepackage{tikz}
\usetikzlibrary{positioning,arrows.meta}
\usepackage[export]{adjustbox}
\usepackage{subcaption}
\usetikzlibrary{automata,arrows,positioning,calc}
\usepackage{booktabs}
\usepackage{lmodern}
\usepackage{bm}
\usepackage[textsize=footnotesize]{todonotes} 
\usepackage[colorlinks=true, linkcolor=blue, citecolor=blue]{hyperref}
\usepackage[numbers]{natbib}
\usepackage{enumitem}
\usepackage{makecell}

\theoremstyle{definition}
\newtheorem{definition}{Definition}[section] 

\title{Modeling dependency between operational risk losses and macroeconomic variables using Hidden Markov Models}
\author{Nikeethan Selvaratnam\textsuperscript{a, b, 1}, Dorinel Bastide\textsuperscript{a}, Clément Fernandes\textsuperscript{c} \\ and  Wojciech Pieczynski\textsuperscript{c}}
\date{April 2025}

\begin{document}
\newgeometry{left=2.5cm,bottom=2cm, top=1cm, right=2.5cm}

\maketitle
\sloppy  

{\let\thefootnote\relax\footnotetext{\textsuperscript{a} \textit{BNP Paribas Stress Testing Methodologies \& Models. This article represents the opinions of the author, and it is not meant to represent the position or opinions of BNP Paribas or its members.}}
}
{\let\thefootnote\relax\footnotetext{\textsuperscript{b} \textit{Laboratoire Services répartis, Architectures, Modélisation, Validation, Administration des Réseaux (Samovar), CNRS UMR 5157, Télécom SudParis, Institut Polytechnique de Paris.}}}

{\let\thefootnote\relax\footnotetext{\textsuperscript{c} \textit{Laboratoire Services répartis, Architectures, Modélisation, Validation, Administration des Réseaux (Samovar), CNRS UMR 5157, Télécom SudParis, Institut Polytechnique de Paris.}}}

{\let\thefootnote\relax\footnotetext{\textsuperscript{1} We gratefully acknowledge Olivier Derollez and Elisa Ndiaye for their constructive feedback and valuable suggestions on earlier drafts of this article.}}
\begin{abstract}
Predicting future operational risk losses gives rise to a significant challenge due to the heterogeneous and time-dependent structures present in real-world data. Furthermore, stress test exercises require examining the relationship with operational losses. To capture such relationship, we propose to use an extension of Hidden Markov Models to multivariate observations. This model introduces a third auxiliary variable designed to accommodate the economic covariates in the time-series data. We detail the unique aspects of operational risk data and describe how model calibration is achieved via the Expectation-Maximization (EM) algorithm. Additionally, we provide the calibration results for the various risk-event types and analyze the relevance of the inclusion of the macroeconomic covariates. 

\end{abstract}


\section{Introduction}
Modeling Operational Risk (OR) has become a major concern for financial institutions in general, holding equal weight alongside credit and market risks. Indeed, instances such as inaccurate financial transactions, system malfunctions or non-compliance with local regulations for financial institutions may result in substantial operational losses and damage their reputation. The evolution in regulatory frameworks, notably through the Basel agreements (\textcolor{red}{\cite{basel2006ii}}, \textcolor{red}{\cite{basel2011iii}}, \textcolor{red}{\cite{basel2017iv}})) shows the ongoing need for improvement in OR management practices. The Basel II agreement precisely defines OR as: \begin{quote} ``the risk of losses due to inadequate or failed internal and external processes, people or environmental events''.
\end{quote} Given the large panel of incidents falling under this scope, a comprehensive mapping by 7 types of incidents and 8 business lines is provided to standardize the computation of capital requirements through all financial institutions that must implement those regulations. Basel II also outlines three approaches for quantification: the Basic Indicator Approach (BIA), the Standardized Approach (SA), and the Advanced Measurement Approach (AMA). The AMA allows banks to use their internal models under strict qualification criteria to quantify and model OR. The global financial crisis of 2007-2008 led to a significant tightening of banking supervision under Basel III, reflecting the need for greater resilience of the banks' balance sheets under stressed financial conditions. Regulators have introduced stress testing exercises (for instance, banking regulations CCAR\footnote{Comprehensive Capital Analysis and Review: regulatory framework introduced by the Federal Reserve for large U.S. banks, designed to assess capital adequacy under severe economic conditions.} or EBA S/T\footnote{The European Banking Authority (EBA) stress test exercise: designed to assess capital adequacy under severe economic conditions for EU financial institutions.}) which emerged as an essential tool in capital planning by evaluating whether the bank remains solvent in times of crisis, and in senior management to make informed decisions about risk appetite, risk mitigation strategies, and capital allocation. Those stress testing exercises rely on simulating macroeconomic shocks and analyzing their propagation through the financial network in a forward looking manner \cite{bastide2023derivatives}, \cite{bellini2016stress}. To ensure comparison between financial institutions applying such regulations and limit the freedom given to internal models (generating too diverse results for operational risk corresponding capital measurements), the introduction of Basel IV in 2023 led to the withdrawal of the AMA for regulatory capital calculation and introduced a revised standardized approach promoting comparability and transparency across institutions.\ However, internal models remain vital for determining economic capital 
which aligns more closely with the institution's actual risk profile and supports various stress-testing exercises as part of the regulatory process SREP\footnote{Supervisory Review and Evaluation Process.} ICAAP reporting.  \newline

\noindent Internal models for measuring operational risk widely use a technique called the Loss Distribution Approach or LDA in which the total operational loss is usually formulated as a random sum \citep[Section 5.1, page 79]{CruzPetersShevchenko2015}. It consists in getting an aggregated loss from a severity distribution (magnitude of the incident) and a frequency distribution (occurrence of the incident). As there are two sources of randomness, this type of process is called a compound process. Consider the total annual loss $L_{t}$ due to operational incidents at a time horizon $t$. In the Basel framework, incidents are split into $C$ categories corresponding to the mapping by event type and business lines. This annual loss can be written as the sum of those $C$ sub-processes:
\begin{equation*}
L_{t}=\sum_{c=1}^{C} L_{t}^{(c)}
\end{equation*}
\noindent where each sub-process $L_{t}^{(c)}$ is a compound process modeling the $c^{th}$ risk cell:


\begin{equation*}
L_{t}^{(c)}=\sum_{m=1}^{N_{t}^{(c)}} S_{t,m}^{(c)}
\end{equation*}

\begin{itemize}
    \item $L_{t}^{(c)}$ is the aggregated loss for the $c^{th}$ risk cell;
    \item $N_{t}^{(c)}$ is a counting process (frequency distribution);
    \item $S_{t,m}^{(c)}$ are identically distributed random variables (severity distribution).
\end{itemize} 

\noindent The Poisson distribution is largely used in the industry to model the frequency of operational risk events as it captures the occurrence of rare events. It assumes that the inter-arrival intervals are independent and have an exponential distribution (lack of memory property), meaning that if an incident does not occur during a given period of time, the probability that it will occur on the next same interval of time does not increase or decrease. The resulting compound process is called a Compound Poisson Process (CPP). Distributions such as the Lognormal, Weibull or Pareto coming from the Extreme Value Theory are usually considered for the loss severity due to their potential to capture the tail behavior of the loss distribution. Once the frequency and severity distributions are calibrated, Monte Carlo simulation techniques are used to get the aggregated loss as it is mostly impossible to get a closed-form formula for the convolution of a random number of non-Gaussian severity distributions. Finally, one can compute the key risk measures. For capital charge, it is the Value-at-Risk or VaR at a $99.9^{th}$ confidence level for a 1-year holding period \citep{BCBS05}. For stress testing purposes, we often consider the $50^{th}$ and $90^{th}$ levels \cite{EBA21, STAMPE2017}. The LDA offers by construction flexibility in modeling the loss as it adapts to the core of the distribution but also to extreme events. Furthermore, this method is granular: institutions can tailor their models by using different distributions for various types of risks, thereby reflecting their specific operational risk profiles and exposures (i.e. the extent to which they are vulnerable to particular risk events). This is actually one of the reasons why it was discarded in the Basel IV framework because of the variability across the models. Nonetheless, due to its possible use for economic capital measurements, the traditional LDA presents several drawbacks that need to be addressed. First of all, estimating both frequency and severity distributions requires extensive historical data especially to capture extreme losses accurately with limited data. Moreover, relying solely on historical data can be problematic as the environment significantly changed in the past years:
emerging risks stemming from plausible cyberattacks or geopolitical events, that were less frequent in the past, should now be taken into account. As a result, the data is complemented with external benchmarks and expert judgement. Using CPP comes with simplified assumptions. The key assumption in the classical Poisson process is that events occur independently and the time between them follows an exponential distribution. In practice, however, operational risk events may be dependent — one shock can trigger multiple losses — and the process often exhibits memory. For example, once a recurring IT system issue is fixed, the likelihood of a similar event in the near future is reduced, contradicting the memoryless property of the Poisson model. CPPs also assume that losses are independent of each other which limits its ability to model correlated risks effectively. Exploring these correlated risks concepts in the LDA framework raised great interest in the literature. For instance, Lindskog and McNeil's Common Poisson Shock Models (\textcolor{red}{\cite{lindskog2003}}) allow for modeling the impact of common causes across processes but are limited to positive correlations. A. Kreinin contributions' (\textcolor{red}{\cite{duch2014new}}, \textcolor{red}{\cite{kreinin2017correlated}}, \textcolor{red}{\cite{chiu2017correlated}}, \textcolor{red}{\cite{bae2017backward}}) extend this approach by incorporating both positive and negative correlations, with a focus on event arrival times. His methodology relies on an algorithm that computes extreme joint distributions, incorporating nonlinear dependencies through Fréchet copulas. Tankov’s work (\textcolor{red}{\cite{tankov2016levy}}, \textcolor{red}{\cite{tankov2019simulation}}) applies Lévy copulas to model dependencies in jumps of Lévy processes, addressing simulation and modeling of simultaneous jumps. Kluppelberg (\textcolor{red}{\cite{boecker2010multivariate}}) expanded this approach by modeling both frequency and severity dependencies using Lévy copulas. However, all of these frameworks pose challenges for calibration, especially if the aim is to dynamically adjust the parameters over time. This time-varying aspect of the framework has received less attention in the literature, and is of great interest in the context of stress testing exercises that require additionally the dependence with economic covariates. The basic CPP does not naturally accommodate time-varying intensities or seasonal effects while operational losses may increase during certain periods (financial crisis, natural disasters or wars). \newline

\noindent Hidden Markov models (HMMs) are a great tool to model this time dependency, especially as operational losses are characterized with regime switches. This idea is not new in operational risk as shown in the following examples. Dionne and Hassani's work (\textcolor{red}{\cite{dionne2016hidden}}) applies HMMs to identify different regimes in loss data, demonstrating how regime switches impacted financial institutions' operational risk during the 2007–2009 crisis. They show that capital estimations can be over or underestimated in periods of normal or more severe losses. While their model accounts for hidden regimes in the data, it does not directly incorporate external economic indicators or other risk factors that could influence operational losses. Fung, Badescu, and Lin have developed in (\textcolor{red}{\cite{fung2019multivariate}}) a multivariate Cox Hidden Markov Model (HMM) to model the joint arrival process of operational risk loss events. This extends the traditional Poisson process by including hidden Markov regimes to account for the dependency between different business lines and event types and combining them with Cox Processes. Using an Expectation-Maximization (EM) algorithm, the model allows for efficient calibration and accurately predicts short-term future losses by capturing the frequency and severity of events across multiple dimensions. Hambuckers and Kneib's research in (\textcolor{red}{\cite{hambuckers2018markov}}) presents a mathematical framework that comes close to what is required in stress testing exercises. More specifically, according to \cite{bellini2016stress}, stress testing should highlight the connection between the overall economy and a given bank: they need to be {\color{blue}both} macroeconomic driven (thus scenario based) and bank-specific (capture how macroeconomic shocks impact capital, liquidity and business model viability). They propose a Markov-Switching Generalized Additive Model (MS-GAM) to account for regime changes in operational risk environments. In this model, CPPs are used to model the arrival of loss events, while the Markov-Switching mechanism captures the transition between different risk regimes over time. The Generalized Additive Model (GAM) grants a flexible structure to model nonlinear relationships between economic covariates and the intensity of the Poisson process, which describes the frequency of losses. The model switches between different states (risk regimes), with each state having its own loss distribution characteristics. These states are governed by a hidden Markov process, which captures unobservable switches in the underlying operational risk environment. Parameters are calibrated using maximum likelihood estimation techniques. This model provides a robust approach to dealing with the complex and regime-dependent nature of operational losses and demonstrates its effectiveness on operational loss data. The relationships between covariates and the losses are estimated nonparametrically using penalized B‑splines. However, unless interaction terms or joint smooth functions are explicitly incorporated, the model may not fully account for dependencies between multiple covariates within and across regimes. \newline

\noindent Unlike traditional applications of Hidden Markov Models (HMMs), which primarily aim to identify hidden states within time series data, our approach focuses on the predictive capabilities of the HMM framework in estimating operational risk quantiles. Rather than using the model to solely infer latent states, we propose to leverage its structure to improve risk estimation by directly incorporating macroeconomic variables into the observation process. Our objective is not only to model the time-varying nature of operational risk, but also to understand whether external financial indicators can help explain or even anticipate changes in the distribution of losses. To this end, we adopt an extension of the standard HMM, generalizing it from univariate to multivariate observation vectors, where one component represents operational losses and the other reflects a macroeconomic covariate. This formulation allows for a direct assessment of whether macroeconomic conditions influence the dynamics of operational loss distributions. By systematically comparing models with and without such covariates across various risk-event categories, we develop a structured methodology to evaluate when their inclusion provides predictive benefits and when, on the contrary, it yields little or no improvement. Our key contribution lies in this flexible modeling framework, which not only accommodates dependence with economic variables but also enables a more granular and scenario-sensitive approach to operational risk estimation. Moreover, by capturing the joint behavior of losses and macroeconomic signals through a latent Markov structure, we aim to reconcile the non-Markovian patterns often observed in operational loss data (such as memory effects, clustering or persistence) with tractability and interpretability of Markovian models. In contrast with traditional approaches such as the Compound Poisson Process, which typically assume stationary and independent structures, our framework offers a more dynamic and flexible perspective, better suited for stress testing exercises.

\noindent 

\section{Methodology}

\subsection{Multivariate model with hidden Markov chain}

With bold notation indicating multidimensional mathematical object, let us consider a Hidden Markov Model $(X_n, \mathbf{Y}_n)_{n\in \mathbb{N^{*}}}$ defined on a probability space $(\Omega, \mathcal{F}, \mathbb{P})$ such that:

\begin{itemize}
    \item  $(X_n)_{n\in \mathbb{N^{*}}}=:X$ is a Markov chain taking its value in a discrete and finite space $ \bigl\{ {\omega_{1}, ..., \omega_{K}} \bigl\}$ with $K>1$. As such, we have for all $n \geq 2$:
    \begin{equation}
        \mathbb{P}[X_{n}=x_{n}|X_{1}=x_{1},..., X_{n-1}=x_{n-1}] = \mathbb{P}[X_{n}=x_{n}|X_{n-1}=x_{n-1}].
    \end{equation}
    The law of $(X_n)_{n\in \mathbb{N^{*}}}$ is given by the initial distribution $\pi$ with $\pi_{i}=\mathbb{P}[X_{1}=\omega_{i}]$ and the transition matrix $\mathbf{A} = \{a_{ij}\}_{1\leq i,j\leq K}$ with $a_{ij} = \mathbb{P}[X_{n}=\omega_{j}| X_{n-1}=\omega_{i}]$, $i,j=1,\dots,K$. As $a_{ij}$ are supposed to be independent of $n$, $(X_n)_{n\in \mathbb{N^{*}}}$ is said to be a homogeneous Markov chain.

    \item $\mathbf{Y}_n$ is a d-dimensional Gaussian random vector. We assume that: (1) given the latent process $(X_n)_{n\in \mathbb{N^{*}}}$, $\mathbf{Y}_n$ is a sequence of conditionally independent random vectors and that (2)  for any $n$, the conditional distribution of $\mathbf{Y}_n$ with respect to $X$ is equal to the conditional distribution of $\mathbf{Y}_n$ with respect to $X_n$ for each $n$ and possesses a density function called the emission probability density function.
    For $i = 1,... ,K$, we denote by $f_{i}( \, \cdot \, ; \bm{\theta}_i)$ the density function of $\mathbf{Y}_n$ conditional to $X_n=\omega_i$ (and that is the same for any $n$), of parameters $\bm{\theta}_{i}$ which belong to a set $\Theta = \mathbb{R}^{d} \times \mathbb{S}^{d}_{++} (\mathbb{R})\footnote{The covariance matrices are symmetric positive definite}$ as it contains the mean and covariance matrix parameters of the Gaussian vector. 
    
\end{itemize}

\noindent As a result, our HMM is fully characterized by the set of parameters denoted by the model $ \Xi = (\pi, \mathbf{A}, \bm{\theta}_{1}, ..., \bm{\theta}_{K})$. Unless otherwise specified, the probability measure $\mathbb{P}$ embeds this model $\Xi$.\newline

\noindent Let $N$ be the length of our observed time series. We denote by $x_{1:N}=(x_1, \dots, x_N)$ and $\mathbf{y}_{1:N}=\mathbf{y}_1, \dots, \mathbf{y}_N$ a realization of length $N$ of the stochastic processes $(X_n)_{n\in \mathbb{N^{*}}}$ and $(\mathbf{Y}_n)_{n\in \mathbb{N^{*}}}$. The joint distribution, which characterizes the dependency between the hidden state sequence $X_{1:N}:=(X_{1},...,X_{N})$ and the corresponding observation sequence $\mathbf{Y}_{1},...,\mathbf{Y_{N}}$ simplifies to (cf. \citep[§III.A, eqs.~(13)--(16)]{rabiner1989}):

\begin{equation}
\begin{aligned}
\mathbb{P}\!\left[X_{1:N}=x_{1:N},\,\mathbf{Y}_{1:N}=\mathbf{y}_{1:N}\right]
&= \mathbb{P}[X_1=x_1]\,
   \mathbb{P}[\mathbf{Y}_1=\mathbf{y}_1\mid X_1=x_1] \\
&\quad \times \prod_{n=2}^{N} \mathbb{P}[X_n=x_n\mid X_{n-1}=x_{n-1}]\,
                        \mathbb{P}[\mathbf{Y}_n=\mathbf{y}_n\mid X_n=x_n].
\end{aligned}
\end{equation}
\noindent As outlined by Rabiner in \cite{rabiner1989}, the practical use of HMMs relies on addressing three core challenges:
\begin{itemize}
    \item \textit{Inference}: Determining the likelihood of an observed sequence given a specific model $\xi$, that is to say we want to compute $\mathbb{P}[\mathbf{Y}_{1:N}=\mathbf{y}_{1:N} \mid \Xi=\xi]$.
    \item \textit{Decoding}: Identifying the most probable sequence of hidden states that could have generated the observed sequence, given the observations $\mathbf{y}$ and the model $\xi$. This is solved using the \underline{Viterbi algorithm}(cf. \citep[§III.D]{rabiner1989}).
    \item \textit{Learning}: Optimizing the model parameters $ \Xi = (\pi, \mathbf{A}, \bm{\theta}_{1}, ..., \bm{\theta}_{K})$ to best fit the given set of observations. This is solved using the \underline{Expectation-Maximization (EM) algorithm}.
\end{itemize}

\noindent In the following, we propose a description of those three algorithms, which will be necessary in our methodological framework. 

\begin{enumerate}[label=\Alph*.]

    \item \underline{The Forward algorithm} \newline
    It is used to efficiently compute the one-step-ahead predictive density in a HMM. Given a model $\Xi=\xi$ and the observations $\textbf{y}_{1:N}$, we want to compute $\mathbb{P}[\mathbf{Y_{1:N}}=\mathbf{y}_{1:N}\mid \Xi=\xi]$. Instead of summing over all possible hidden state sequence (which is exponential in $n$), we use dynamic programming via the Forward algorithm. We introduce for that matter the forward variables $\alpha_{n}(j)$ (cf. \citep[§III.A, eqs.~(18)]{rabiner1989}):

    \begin{definition}[Forward probabilities] 
    It is defined as the probability density function of the partial observations $\mathbf{Y}_{1:n}$ until the state $X_{n}$ given the model $\Xi = \xi$:
    \begin{equation}\label{e:AlphaExpr}
    \alpha_{n}(j) = \mathbb{P}\left[ X_n=\omega_{j}, \mathbf{Y}_{1:n}=\mathbf{y}_{1:n}\mid \Xi = \xi \right] \hspace{1em}, \hspace{0.5em} 1 \leq n \leq N-1 ,
    \end{equation}
    and computed recursively as follows (cf. \citep[§III.A, eqs.~(19)--(21)]{rabiner1989}): 
    \begin{enumerate}
        \item initialization: $\alpha_1(j) = \pi_{j}f_{j}(\mathbf{y}_1; \bm{\theta}_j),\quad  \text{for} \hspace{0.25em} 1\leq j\leq K$, 
        \item induction: $\alpha_{n+1}(j) = \left[ \sum_{k=1}^{K}\alpha_{n}(k)a_{kj}   \right]f_{j}(\mathbf{y}_{n+1}; \bm{\theta}_{j}), \quad \text{for any} \hspace{0.25em} 1\leq n\leq N-1 \hspace{0.25em} \quad  \text{and} \hspace{0.25em}  1 \leq j \leq K $,
        \item termination: $\mathbb{P}[\mathbf{Y}_{1:N}=\mathbf{y}_{1:N} \mid \Xi=\xi] = \sum_{k=1}^{K}\alpha_{N}(k)$ (which corresponds to the computation of the likelihood).
    \end{enumerate}
    \end{definition}

    \noindent From the forward variables, we obtain the one-step ahead predictive density which is given by:
    \begin{equation}
    \mathbb{P}\!\left(\mathbf{Y}_{n+1}=\mathbf{y}_{n+1}\,\middle|\,\mathbf{Y}_{1:n}=\mathbf{y}_{1:n}\right)
    = \sum_{k=1}^{K}\sum_{j=1}^{K} 
    \underbrace{\mathbb{P}(X_n=\omega_k \mid \mathbf{Y}_{1:n}=\mathbf{y}_{1:n})}_{\displaystyle \alpha_n(k)/\sum_{\ell=1}^{K}\alpha_n(\ell)}\,
    a_{kj}\, f_j(\mathbf{y}_{n+1};\bm{\theta}_j).
    \end{equation}

    \noindent Note that the normalized version $\alpha^{\star}_n(k)=\frac{\alpha_n(k)}{\sum_{\ell=1}^{K}\alpha_n(\ell)}$ is called the filtered belief state or filtered posterior at time $n$: the forward algorithm allows you to carry this belief and predict what you are likely to observe next.
    
    \item \underline{The Forward-Backward algorithm} \newline
    We denote by $\phi_{n}(i)$ the smoothed posterior marginal:
    \begin{align}\label{e:phiExpr}
        \phi_{n}(i) = \mathbb{P}[X_{n }=\omega_{i}\mid \mathbf{Y}_{1:N}=\mathbf{y}_{1:N} , \Xi=\xi]
    \end{align}

    Note that this smoothed posterior marginal should not be confused with the filtered posterior distribution $\mathbb{P}(X_n=\omega_i \mid \mathbf{Y}_{1:n}=\mathbf{y}_{1:n})$, which only depends on past and present observations. In contrast, $\phi_n(i)$ also incorporates the future observations (from $n+1$ to $N$). This smoothing distribution is mainly used in the parameter estimation step (Section~C, Baum--Welch algorithm) and for decoding.
    To simplify the computation of this quantity, we introduce the backward probabilities $\beta_{n}(j)$ (cf. \citep[§III.A, eqs.~(23)]{rabiner1989}):

    \begin{definition}[Backward probabilities] It is defined as the probability density function of the partial observations $\mathbf{Y}_{n+1:N}$ given that we are in the state $X_{n}=\omega_i$ and the model $\Xi$:
    \begin{align}\label{e:BetaExpr}
    \beta_{n}(i)= \mathbb{P} \left[ \mathbf{Y}_{n+1:N}=\mathbf{y}_{n+1:N} \mid X_n=\omega_{i}, \Xi=\xi \right] \hspace{1em} \text{for} \hspace{0.5em} 1 \leq n \leq N-1,
    \end{align}
    and computed recursively as follows: 
    \begin{enumerate}
        \item initialization: $\beta_N(i) = 1, \text{for} \hspace{0.25em} 1 \leq i \leq K$,
        \item induction: $\beta_{n}(i) =  \sum_{j=1}^{K}a_{ij}f_{j}(\mathbf{y}_{n+1}; \bm{\theta}_{j}) \beta_{n+1}(j), \text{for any} \hspace{0.25em} 1 \leq n \leq N-1 \hspace{0.25em}  \text{and} \hspace{0.25em}  1 \leq  \leq K $,
        \item termination: $\mathcal{L}(\mathbf{y}\mid \Xi) := \sum_{k=1}^{K}\beta_{1}(k) \pi_{k}f_{k}(\mathbf{y}_{1}; \bm{\theta}_{k})$.
    \end{enumerate}
\end{definition}

    The vector $\mathbf{\beta}_{n}(i)$ of size $n \times K$ corresponds to the conditional likelihood of future evidence given the hidden state at step $n$. 

    Finally, the smoothed posterior marginal can be decomposed into the past and future components by conditioning on the belief state $x_{n}$ (cf. \citep[§III.B, eqs.~(27)]{rabiner1989}):
    \begin{align}\label{e:XiExpr}
        \phi_{n}(i) = \frac{\alpha_{n}(i)\beta_{n}(i)}{\sum_{k=1}^{K}\alpha_{n}(k)\beta_{n}(k)}
    \end{align}

    \noindent As both the forward and backward variables naturally decrease with each iteration of the algorithm, they eventually reach values so small that they are numerically rounded to zero, leading to potential underflow issues. (cf. \citep[§V.A, eqs.~(93b)--(94)]{rabiner1989}) To prevent these numerical instabilities, it is necessary to implement an appropriate scaling procedure at each iteration. Specifically, during the forward step, we define the scaling factors

    \begin{equation}
         s_n = \frac{1}{\sum_{k=1}^K \alpha_n(k)}
    \end{equation}
    \begin{equation}
          \alpha^{\star}_n(j) = s_n\, \alpha_n(j)   \hspace{0.5em} \text{and} \hspace{0.5em}\beta^{\star}_n(j) = s_n\, \beta_n(j)
    \end{equation}

    Intuitively, the forward-backward algorithm propagates information first from left to right and then from right to left, integrating both directions at each time step. While each procedure can independently compute the likelihood of the observation sequence, both are required jointly to identify the model $\Xi$ parameters that maximize this likelihood.



    \item \underline{The Baum-Welch algorithm} \newline
     Because the sequence of states followed by the Markov chain in a Hidden Markov Model is unobserved, it is standard to treat these states as latent variables and apply the Expectation-Maximization (EM) algorithm, known as the Baum-Welch algorithm in the case of HMMs ((cf. \citep[Ch.~11]{cappe2005inference}) for a precise derivation). Given the sequence of observations $\mathbf{Y}_{1:N}$, the aim is to find by maximum likelihood estimation:
     \begin{align}
         \underset{\xi}{\arg\max} \mathbb{P}[\mathbf{Y}_{1:N}=\mathbf{y}_{1:N}|\Xi=\xi] = \underset{\xi}{\arg\max} \sum_{x_{1:N}\in \{1,\dots,K\}^N} \mathbb{P}[
         X_{1:N}=x_{1:N},\mathbf{Y}_{1:N}=\mathbf{y}_{1:N}|\Xi=\xi]
     \end{align}

    \noindent The algorithm relies on having both the forward and backward probabilities, which are obtained through the forward-backward procedure.
    \noindent In the Expectation-Maximization (EM) algorithm, we aim to maximize the marginal log-likelihood of the observed data by iteratively estimating the hidden variables and updating the model parameters $ \Xi = (\pi, A, \bm{\theta}_{1}, ..., \bm{\theta}_{K})$.  Note that $\bm{\theta}_{i}=(\bm{\mu}_{i},\bm{\Sigma}_{i}) \hspace{0.5em}\text{for} \hspace{0.5em} 1\leq i\leq K$ as in our framework, we have Gaussian emissions. We introduce the following variables that help us compute the complete-data log-likelihood in the Expectation-step and derive closed-form updates for the parameters  of the model:
    \begin{itemize}
    \item $ \phi_n(i) = \mathbb{P}(X_n = \omega_i \mid \mathbf{Y}_{1:N}=\mathbf{y}_{1:N}, \Xi=\xi)$: the posterior probability that the hidden state at time $n$ is $\omega_{i}$, given the full observation sequence, as defined earlier in Equation \ref{e:phiExpr}.
    \item $\Psi_{n}(i,j) = \mathbb{P}(X_n = \omega_i, X_{n+1} = \omega_j \mid \mathbf{Y}_{1:N}=\mathbf{y}_{1:N}, \Xi=\xi)$: which is the probability of being in state $\omega_i$ at time $n$, in state $\omega_j$ at time $n+1$, given the model and the observation sequence
    \end{itemize}
    These two quantities can be expressed using the forward and backward variables defined in \eqref{e:AlphaExpr} and \eqref{e:BetaExpr} using the Bayes' rule for a given model $\xi$. For $\phi_n(i)$, we have: 
    \begin{align}
        \phi_n(i) = \mathbb{P}(X_n = \omega_i \mid \mathbf{Y}_{1:N}=\mathbf{y}_{1:N}, \Xi=\xi) = \frac{\alpha_n(i) \beta_n(i)}{\sum_{k=1}^K \alpha_n(k) \beta_n(k)},
    \end{align}
and for $\Psi_n(i, j)$ (cf. \citep[§III.C, eqs.~(37)]{rabiner1989}):

\begin{align}
\Psi_n(i, j)
&= \frac{\alpha_n(i)\, a_{ij}\, f_{j}(y_{n+1} ; \bm{\mu}_j, \bm{\Sigma}_j)\, \beta_{n+1}(j)}{
\sum_{i=1}^{K} \sum_{j=1}^{K} \alpha_n(i)\, a_{ij}\, f_{j}(y_{n+1} ; \bm{\mu}_j, \bm{\Sigma}_j)\, \beta_{n+1}(j)}
\end{align}



The algorithm is derived as follows:
\begin{enumerate}
    \item \textbf{E-Step}: Computation of the forward and backward recursions. The quantities $\phi_{n}(i)$ and $\Psi_n(i,j)$ are calculated given the formulas above.
    \item \textbf{M-Step}: Parameters updates. It is a constrained optimization problem where the log likelihood of the observations sequence is maximized, which leads to the following expressions\footnote{In practice, one may add a small ridge term $\epsilon \in \mathbb{I}$ to ensure that the covariance matrix is positive definite \citealp{bishop2006pattern} }:
    \begin{equation}
    \begin{aligned}
    \pi_{i}^{\text{new}} &= \phi_1(i), 
    &\quad 
    a_{ij}^{\text{new}} &= \frac{\sum_{n=1}^{N-1} \Psi_n(i, j)}{\sum_{n=1}^{N-1} \phi_n(i)}, \\[0.5em]
    \bm{\mu}_i^{\text{new}} &= \frac{\sum_{n=1}^N \phi_n(i)\, \mathbf{y}_n}{\sum_{n=1}^N \phi_n(i)}, 
    &\quad 
    \bm{\Sigma}_i^{\text{new}} &= \frac{\sum_{n=1}^N \phi_n(i)\, (\mathbf{y}_n - \bm{\mu}_i^{\text{new}})(\mathbf{y}_n - \bm{\mu}_i^{\text{new}})^\top}{\sum_{n=1}^N \phi_n(i)} 
    \end{aligned}
    \end{equation}
   
    The log-likelihood of the observations is also given by, using the scaled forward probabilities as recommended numerically: \begin{align}\label{e:LLExpression}
    \log P(\mathbf{Y}_{1:N}{=\textbf{y}_{1:N}}\mid \Xi=\xi) = - \sum_{n=1}^{N} \log s_n
    \end{align}
    
\end{enumerate}
The EM algorithm requires an initial estimation of the model parameters, which can significantly affect both the convergence speed and the quality of the final solution. A common practice is to initialize the emission parameters $\bm{\theta}_i$ using K-means clustering on the observation sequence $\mathbf{Y}_{1:N}$, assigning each cluster to a hidden state. This initialization provides a partition of the data in the absence of prior knowledge and helps avoid poor local optima during the early iterations of EM \cite{bishop2006pattern}. However, the EM algorithm is only guaranteed to converge to a local maximum of the likelihood function, not the global maximum. Therefore, the choice of initialization can potentially lead to convergence towards a suboptimal solution far from the global maximum \cite{murphy2012machine, bishop2006pattern}. To mitigate this risk, it is common to run the EM algorithm multiple times with different random or data-driven initializations and retain the solution that gives the highest likelihood.

\end{enumerate}

\subsection{Application to Operational Risk Modeling}

We consider a multivariate HMM where the hidden states $ X_n \in \{\omega_1, \dots, \omega_K\} $ form a discrete-time Markov chain, and the observations  $ (Y_n^1, Y_n^2)_{n \in \mathbb{N}}$ takes its values in  $\mathbb{R}^2 $ and follows a Gaussian distribution conditional on the latent state:
\begin{align*}
(Y_n^1, Y_n^2) \mid \{ X_n = \omega_i \}\sim \mathcal{N}(\bm{\mu}_i, \bm{\Sigma}_i), \quad i = 1, \dots, K,
\end{align*}
with $\bm{\theta}_i=(\bm{\mu}_i, \bm{\Sigma}_i)$.

\noindent We also assume that for each $i=1,\dots,K$, the parameters have been already calibrated using the Baum-Welch algorithm described in the previous section and as such the considered probability measure $\mathbb{P}$ is conditional to this model calibration that is $\mathbb{P}(\,\cdot\,)=\mathbb{P}(\,\cdot\,|\Xi=\xi)$ with $\xi=(\pi,\mathbf{A},\bm{\theta}_1,\dots,\bm{\theta}_K)$.

\begin{figure}[h!]
\centering
\begin{tikzpicture}[every node/.style={circle, draw, minimum size=0.7cm, inner sep=0pt}, scale=1.2]

\node (x1) at (0,0) {\(x_1\)};
\node (x2) at (2,0) {\(x_2\)};
\node (xN) at (6,0) {\(x_N\)};
\node (xNp1) at (8,0) {\(x_{N+1}\)};

\node (y11) at (0,1.5) {\(y_1^1\)};
\node (y21) at (2,1.5) {\(y_2^1\)};
\node (yN1) at (6,1.5) {\(y_N^1\)};
\node (yNp11) at (8,1.5) {\(y_{N+1}^1\)};

\node (y12) at (0,3) {\(y_1^2\)};
\node (y22) at (2,3) {\(y_2^2\)};
\node (yN2) at (6,3) {\(y_N^2\)};
\node (yNp12) at (8,3) {\(y_{N+1}^2\)};

\draw (x1) -- (x2);
\draw[dotted] (x2) -- (xN);
\draw (xN) -- (xNp1);

\draw (x1) -- (y11);
\draw (x2) -- (y21);
\draw (xN) -- (yN1);
\draw (xNp1) -- (yNp11);

\draw[bend left=25] (x1) to (y12);
\draw[bend left=25] (x2) to (y22);
\draw[bend left=25] (xN) to (yN2);
\draw[bend left=25] (xNp1) to (yNp12);

\draw (y11) -- (y12);
\draw (y21) -- (y22);
\draw (yN1) -- (yN2);
\draw (yNp11) -- (yNp12);

\end{tikzpicture}
\caption{Graphical model of a bivariate Gaussian HMM - the discrete latent variables $x_n$ introduce dependency between the operational losses $Y_{n}^{2}$ and the macroeconomic variable $Y_n^{1}$}
\label{bivariate-hmm}
\end{figure}
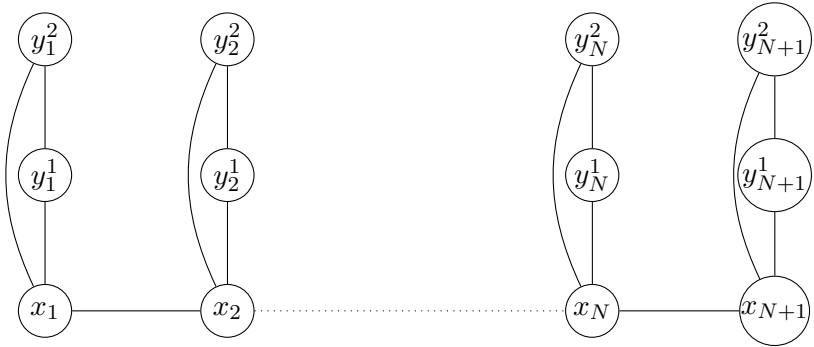

\noindent This setting corresponds to the bivariate ($d=2$) case of the model exposed in Section 2.1. The first component $Y_n^1$ represents a macroeconomic or financial covariate and the second component $Y_n^2$ corresponds to the operational loss variable. In this framework, the latent state $X_{n}$ acts as a source of dependence between the two observable processes, which now exhibits temporal dependence due to the Markovian structure of the latent variable. The use of a hidden state allows us to capture the regimes (e.g. stress vs normal) that simultaneously affect both the losses and the macroeconomic indicators. Figure~\ref{bivariate-hmm} illustrates this bivariate HMM structure and how the dependence between the losses and the macroeconomic variable is modeled.
Such use of a hidden state is also relevant in stress testing contexts where we aim to assess how macroeconomic shocks propagate to operational loss distributions. \newline

\noindent More precisely, we are interested in estimating the conditional $90^{th}$ quantile of the operational losses at a future time step $N+1$, denoted $\varepsilon_{N+1}$, given the full history of past observations:

\begin{equation}
    \varepsilon_{N+1}:=\inf\left\{
    \varepsilon \,\middle|\, \mathbb{P}\left(Y_{N+1}^2 \geq \varepsilon \mid Y_{1:N}^1=y^1_{1:N}, Y_{1:N}^2=y^2_{1:N} \right) \leq 0.1
    \right\}
\end{equation}

\noindent To estimate this quantile, we simulate $M$ independent realizations of $Y_{N+1}^2$  from the conditional predictive distribution using the inferred HMM parameters:

\begin{equation}
    Y_{N+1}^2 \sim \mathbb{P}\left(Y_{N+1}^2 \,\middle|\, Y_{1:N}^1, Y_{1:N}^2\right),
\end{equation}

\noindent The samples are sorted:
\begin{equation}
    Y_{N+1}^{2,(1)} \leq Y_{N+1}^{2,(2)} \leq \dots \leq Y_{N+1}^{2,(M)}
\end{equation}

\noindent and the estimator $ \widehat{\varepsilon}_{N+1} $ is defined as:

\begin{equation}
    \widehat{\varepsilon}_{N+1} = Y_{N+1}^{2,(i)} \quad \text{such that} \quad Y_{N+1}^{2,(i-1)} \leq 0.9M, \quad Y_{N+1}^{2,(i)} \geq 0.9M. 
\end{equation}

\noindent The samples can be generated using the one-step-ahead predictive density distribution, given by:

\begin{equation}
\begin{aligned}
\mathbb{P}\!\left(Y_{N+1}^2 = y_{N+1}^2 \mid Y_{1:N}^1=y_{1:N}^1,\, Y_{1:N}^2=y_{1:N}^2\right)
&= \sum_{i}\;
   \underbrace{\mathbb{P}\!\left(X_N=\omega_i \mid Y_{1:N}^1=y_{1:N}^1,\, Y_{1:N}^2=y_{1:N}^2\right)}_{\frac{\alpha_N(i)}{\sum_{\ell=1}^K \alpha_N(\ell)}} \\
&\quad \times \sum_{j}\;
   \underbrace{\mathbb{P}\!\left(X_{N+1}=\omega_j \mid X_N=\omega_i\right)}_{a_{i,j}} \\
&\quad \times
   \underbrace{\mathbb{P}\!\left(Y_{N+1}^2=y_{N+1}^2 \mid X_{N+1}=\omega_j\right)}_{f_{2}(y_{N+1}^2 \,; \bm{\theta}_j)} 
\end{aligned}
\end{equation}

\noindent where $\frac{\alpha_N(i)}{\sum_{\ell=1}^K \alpha_N(\ell)} = \mathbb{P}(X_N=\omega_i \mid Y_{1:N}^1= y_{1:N}^1,\, Y_{1:N}^2= y_{1:N}^2)$ are the filtered posterior probabilities at time $N$ obtained from the Forward algorithm, $a_{i,j}$ are the transition probabilities, and $f_{2}(\cdot;\bm{\theta}_j)$ denotes the emission density of the second component $Y^{2}$ under the state $\omega_j$.

\noindent In practice, the simulation of $Y_{N+1}^2$ proceeds as follows: 
\begin{enumerate}
    \item \underline{Calibration}: estimate the model parameters 
    $\Xi=(\pi,\mathbf{A},\bm{\theta}_1,\dots,\bm{\theta}_K)$ with the Baum–Welch algorithm. 
    \item \underline{Filtering}: compute the filtered posterior 
    $\mathbb{P}(X_N=\omega_i \mid Y_{1:N}^1=y_{1:N}^1, Y_{1:N}^2=y_{1:N}^2)$ using the forward recursion. 
    \item \underline{State propagation}: draw $X_{N+1}$ from 
    $\mathbb{P}(X_{N+1}=\omega_j \mid X_N=\omega_i)=a_{ij}$. 
    \item \underline{Emission}: sample $Y_{N+1}^2$ from the Gaussian density 
    $f_{2}(\cdot;\bm{\theta}_j)$ linked to $X_{N+1}$. 
\end{enumerate}

\subsection{Choice of the parameters}

\noindent The performance of an HMM depends largely on the selection of appropriate parameters that describe best the hidden state and the link to the observed data. The key parameters are:

\begin{enumerate}
    \item Number of Hidden states $k$. We consider that $X$ is a discrete random variable which takes its values in $\Omega =\{2, ..., 6\}$. Considering too many classes can lead to overfitting. The state space of $X$ can be chosen on specific knowledge about the data or statistical model selection criteria such as the Akaike Information Criterion (AIC) or the Bayesian Information Criterion (BIC).
    \item Transition Matrix $\mathbf{A} = \{a_{ij}\}_{1\leq i,j\leq K}$ that gives the probability of going from one hidden state $i$ to another $j$. In real-world systems, hidden states often exhibit temporal persistence, which means that $a_{ii}$ should be relatively high. 
    \item Initial distribution $\pi$: it is essential to properly quantify it for the forward-backward algorithm, mainly based on prior knowledge about the system.
    \item Emission probabilities $\mathbb{P}  \left[\mathbf{Y}=\mathbf{y} \mid X_n\in E     \right], E\in\mathcal{F}$: we  formulate these quantities based on a multivariate Gaussian distribution.
\end{enumerate}

\noindent Gaussian distributions are often used in HMMs as many real-world processes can be well-approximated by them (see \citep[§III.B, eq.~(18)]{rabiner1989,cappe2005inference}). Furthermore, they are mathematically convenient because they are fully characterized by two parameters: the mean and the covariance matrix. This simplicity allows for efficient parameter estimation using techniques such as the Expectation-Maximization (EM) algorithm (the functions are smooth and differentiable). Through the covariance matrix, we can capture correlations between different dimensions of the data (in our case between the observed loss and the macroeconomic variable). Finally, in the case where a single Gaussian variable is not sufficient to capture the complexity of the data distribution, we can extend the model to Gaussian Mixtures Models (GMM). Using GMM, one can approximate any continuous probability distribution to a selected level of accuracy (see \citep[Ch.~9, §9.2.2]{cappe2005inference,mclachlan2000finite}), enabling complex data modeling. \newline

\noindent  For the economic variable , we use the VSTOXX, which measures market expectations of near-term volatility based on S{\&}P 500 index options\footnote{see for instance
\url{https://stoxx.com/index/v2tx/}
}. It is relevant in models because it captures investor sentiment, specifically their outlook on future market risk and uncertainty. The VSTOXX is often used in financial models for forecasting volatility, managing risk, and pricing derivatives. Since volatility is a key driver of market dynamics, incorporating the VSTOXX could help detect regime switches and have better estimates of the losses. 

\subsection{Risk measure}

Back-testing techniques are used to assess the accuracy and reliability of the different models by comparing their predictions against actual historical data \citep{Jorion2011}. More precisely, we will use the Value-at-Risk (VaR) back-testing at $90\%$ confidence level which is a reference for stress testing exercises \citep{EBA21}.
The concept is to assess whether the actual losses exceed the predicted VaR at a given confidence level over a certain time horizon. In our case, we predict a 90\% VaR, thus actual losses should exceed this threshold approximately 10\% of the time. Such exceedance is known as an exception. The main objective is to compare the number of exceptions with the expected number under the assumption that the model is correct. Let $\text{VaR}_{t}^{\alpha}$ be the $\alpha$-level Value-at-Risk at time $t$ and $L_{t}$ the actual loss at $t$. We define the following function for exceptions:
$$
I_{t} =  \left\lbrace \begin{matrix}
1 & \text{if } L_{t} \geq \text{VaR}_{t}^{\alpha} \\
0 & \text{otherwise.}
\end{matrix}
\right.
$$

\noindent The exception rate given a total number of observations $T$ is given by:

$$
\text{Exception rate} = \frac{1}{T} \sum_{t=1}^{T} I_{t}
$$

\noindent If the number of exceptions exceeds expectations, it suggests that the model may underestimate the level of risk. However, if the number of exceptions is lower than expected, the model might be too cautious or overly conservative. Back-testing in stress testing can be challenging as the stress scenarios are hypothetical and could never have occurred in the past. Nonetheless, one can still rely on the past crisis to simulate similar extreme events.

\section{Empirical Study}

In this section, we apply the proposed model to a dataset of operational losses inspired by some rescaled bank losses, classified according to the seven event-type categories defined in the Basel II regulatory framework (see \cite[Chapter 14, page 499]{CrouhyGalaiMark2014}, \cite[Section 1.4, page 10]{CruzPetersShevchenko2015}). These categories cover a broad spectrum of operational risks faced by financial institutions, as detailed in Table \ref{tab:basel_categories}. 

\begin{table}[!ht]
    \centering
    \renewcommand{\arraystretch}{1.2}
    \setlength{\tabcolsep}{6pt} 
    \begin{tabular}{p{6cm} p{10cm}} 
        \toprule
        \textbf{Category} & \textbf{Description} \\
        \midrule
        Internal Fraud & Unauthorized trading, embezzlement, intentional misreporting. \\
        External Fraud & Theft, cyber fraud, third-party vendor fraud. \\
        Employment Practices & Discrimination, workplace safety violations, wrongful termination. \\
        Clients, Products \& Business Practices & Breach of fiduciary duty, misselling, privacy violations. \\
        Damage to Physical Assets & Natural disasters, terrorism, vandalism. \\
        Business Disruption \& System Failures & IT failures, cyberattacks, telecommunication outages. \\
        Execution, Delivery \& Process Management & Transaction errors, incorrect data entry, settlement failures. \\
        \bottomrule
    \end{tabular}
    \caption{Basel II Operational Risk Event-Type Categories}
    \label{tab:basel_categories}
\end{table}

\noindent While we apply our methodology to all seven categories, we focus in detail on the Internal Fraud category. This choice is motivated by its substantial contribution to overall operational losses in banks and by its significant regulatory and reputational implications. The Basel Committee (BCBS) has repeatedly emphasized the importance of strengthening internal control mechanisms to mitigate risks associated with fraudulent activities.
Empirical studies further support this focus: \citet{chernobai2011operational} and \citet{cope2012macroenvironmental} show that internal fraud losses often exhibit heavy-tailed characteristics and can have severe financial consequences. More precisely, this category refers to cases where bank employees engage in fraudulent activities for personal gain or to manipulate financial statements. According to Basel II, this includes unauthorized trading, such as the Société Générale case with Jérôme Kerviel in 2008, which led to a $4.9$ billion Euro loss \cite[Chapter 14, page 499]{CrouhyGalaiMark2014}.
It also covers theft of assets or embezzlement, where employees transfer funds into personal accounts, falsification of financial transactions, such as manipulation of loan documents, and bribery or corruption, where financial decisions are influenced by personal interests rather than proper risk assessment. These types of fraud are often linked to poor governance and internal controls \cite{hess2011impact, fiordelisi2014determinants}. Studies suggest that financial crisis increases the likelihood of such behavior, as employees might try to cover up losses or take advantage of regulatory loopholes. \newline

\noindent The data set considered for this empirical study consists of losses that mimic certain operational losses that could have been recorded by a large financial institution such as a major bank between January 2000 and June 2018. Due to confidentiality aspects, losses have been altered and anonymized but the interesting patterns have been kept. Each recorded loss reflects a specific operational event, categorized according to the Basel II event taxonomy. To account for financial-market uncertainty, we include the EURO STOXX 50 Volatility Index (VSTOXX; ticker V2TX) as a macroeconomic variable. VSTOXX measures near-term implied equity volatility from EURO STOXX 50 option prices\footnote{See \url{https://stoxx.com/index/v2tx/}.} and is widely used as a proxy for market uncertainty and systemic risk. Prior work \citep{ChavezDemoulin2016} shows that periods of high volatility affect operational loss distributions, particularly for categories such as internal fraud. Incorporating VSTOXX could help the model capture shifts in market conditions (regime changes) and improve the estimation of losses. The main objective of this study is to examine whether the proposed model can identify and quantify a statistical dependence between operational losses and macroeconomic variables. Specifically, we investigate whether incorporating macro-financial variables such as the VSTOXX help capture the dynamics in internal fraud losses. By estimating the 90$^{\text{th}}$ quantile of $L_t$, we aim to determine whether periods of high market volatility are associated with distinct loss regimes, thus providing evidence of a link between financial-market uncertainty and operational risk dynamics for this category. \newline

\noindent The Gaussian framework, although convenient for its analytical tractability, can be difficult to estimate. Indeed, extreme events present in the data can significantly alter parameters estimation and bias the assessment of risk. To reduce this influence and to be able to calibrate the Gaussian vector, a filtering procedure based on quantile thresholds is applied to remove observations outside a predefined range. Given a dataset of operational losses $\{Y_i\}_{i=1}^{T}$, we compute the first and third quartiles, denoted $Q_1$ and $Q_3$, and define the inter-quartile range (IQR) as $IQR = Q_3 - Q_1$. Extreme values are excluded when they fall outside the following bounds \citep{tukey1977eda}:
\begin{equation}
Q_1 - 1.5 \times IQR \leq X_i \leq Q_3 + 1.5 \times IQR.
\end{equation}

\noindent The simplified setup also allows for a first glance at the model's dependence structure, before extending the framework to heavy-tailed elliptical distributions, which may better capture the empirical properties of operational losses.

\subsection{Experimental Plan}

\noindent This experimental analysis investigates whether the inclusion of a macroeconomic volatility indicator improves the loss predictions for operational risk, in particular, whether it helps capture a statistical dependence between losses and macroeconomic conditions. We compare models that differ along three controlled factors: (i) the time aggregation of losses (weekly, monthly, quarterly), (ii) the number of hidden states $K \in \{2,3,4\}$, and (iii) the inclusion of a macroeconomic covariate (VSTOXX). The main performance measure is the one-step-ahead prediction of the 90\% loss quantile, compared to the realized loss. The experimental grid is designed to examine how model specification choices affect the detection of signals between macroeconomic covariates and losses. Varying the time aggregation of losses allows us to capture dependencies that may emerge at different temporal scales. Indeed, a finer frequency (e.g., weekly) captures short-term movements and isolated loss events, but may also introduce noise. In contrast, aggregating over longer horizons (e.g., quarterly) smooths temporal fluctuations and helps reveal broader trends or persistent risk regimes. In operational risk, this aspect is important since losses can be irregular or clustered. We perform this analysis as a similar trade-off appears in image segmentation. In that context, increasing the granularity of the data, that is to say using a finer pixel resolution, enhances the detection of local details at the cost of adding more noise leading to unstable classification in some cases. On the other hand, with a higher resolution, we can identify broader regions in the image \cite{zhang2001segmentation}. \newline

\noindent We estimate HMMs using three levels of temporal aggregation for the loss data: quarterly, monthly, and weekly. The number of hidden states $K$ is varies from 2 to 4. Models with more states were also tested (up to 6) but are not reported here for clarity, as they did not improve interpretability. The EURO STOXX~50 implied volatility index (VSTOXX) is used as the macroeconomic variable to test whether including market volatility can help explain variations in operational losses. The different model configurations are summarized in Table~\ref{tab:configs}. The following pattern is used to identify each model: \texttt{[Aggregation]-[K]-HMM} when the macroeconomic variable is not included and \texttt{[Aggregation]-[K]-M} when it is. For instance, the quarterly three-state HMM is denoted \texttt{Q-3} whereas the one including the macroeconomic variable is \texttt{Q-3-M}. \newline

\begin{table}[h!]
\centering
\small
\renewcommand{\arraystretch}{1.05}
\setlength{\tabcolsep}{6pt}
\begin{tabular}{lcc}
\toprule
\textbf{Aggregation} & \textbf{Hidden states ($K$)} & \textbf{Model labels} \\
\midrule
Quarterly & 2, 3, 4 & \texttt{Q-2 / Q-2-M}, \quad \texttt{Q-3 / Q-3-M}, \quad \texttt{Q-4 / Q-4-M} \\
Monthly   & 2, 3, 4 & \texttt{M-2 / M-2-M}, \quad \texttt{M-3 / M-3-M}, \quad \texttt{M-4 / M-4-M} \\
Weekly    & 2, 3, 4 & \texttt{W-2 / W-2-M}, \quad \texttt{W-3 / W-3-M}, \quad \texttt{W-4 / W-4-M} \\
\bottomrule
\end{tabular}
\caption{Model configurations by aggregation level, number of hidden states, and inclusion of the macroeconomic variable}
\label{tab:configs}
\end{table}

\noindent For each model configuration, the parameters are estimated as described in Section~2.1.C via the EM algorithm on the whole dataset. After model estimation, at each time step, we draw $1{,}500$ samples from the predictive density defined in Equation~(19), and derive the empirical $90^{th}$ quantile. This procedure is repeated recursively to generate the full path of predictive quantiles. Model performance is assessed using backtesting measures described in Section~3.2 by computing the mean squared error of exceedance.

\subsection{Numerical Results on the Internal Fraud Category}

\begin{figure}[h]
    \centering
        \includegraphics[width=1\linewidth]{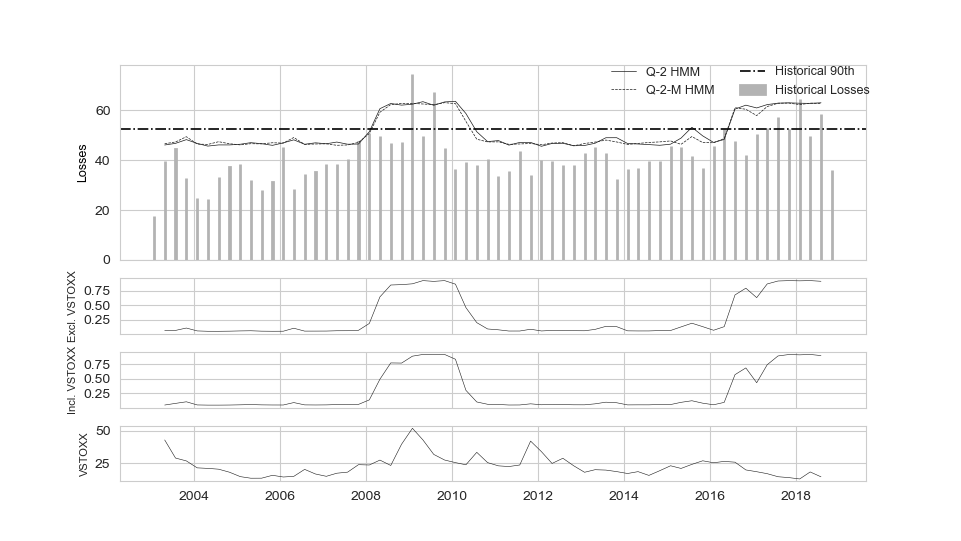}
        \caption{Quarterly 90th–percentile predictions for 2-state HMMs with and without the VSTOXX covariate. grey bars = realized quarterly operational losses; solid line = Q-2 HMM; dashed line = Q-2-M HMM (with VSTOXX); dash–dot horizontal line = historical 90th percentile of losses. Middle: filtered probability of the high-loss state for the models excluding and including VSTOXX, respectively; Bottom: VSTOXX covariate.}
    \label{fig:HMM_2_Q}
\end{figure}

\begin{figure}[h]
    \centering
        \includegraphics[width=1\linewidth]{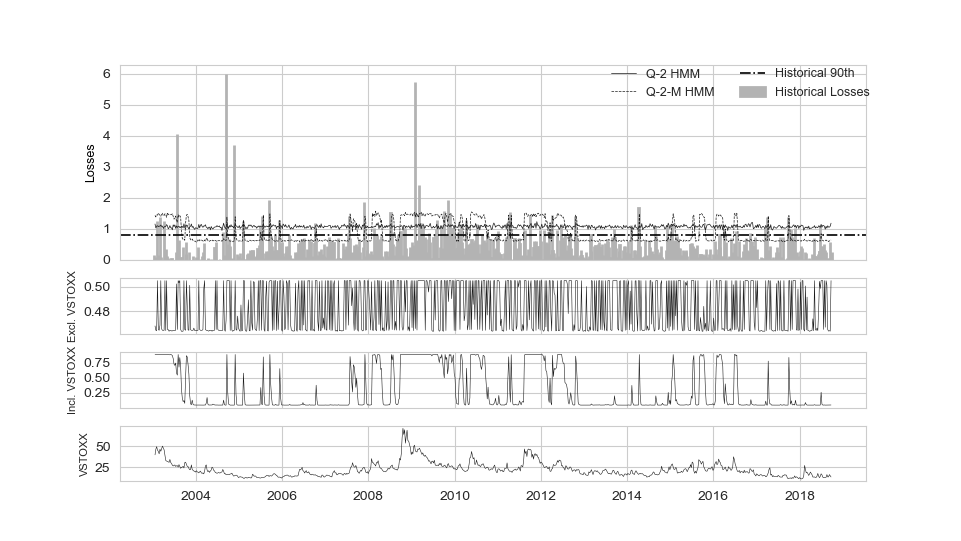}
        \caption{Weekly 90th–percentile predictions for 2-state HMMs with and without the VSTOXX covariate. grey bars = realized weekly operational losses; solid line = Q-2 HMM; dashed line = Q-2-M HMM (with VSTOXX); dash–dot horizontal line = historical 90th percentile of losses. Middle: filtered probability of the high-loss state for the models excluding and including VSTOXX, respectively; Bottom: VSTOXX covariate}
    \label{fig:HMM_2_W}
\end{figure}

Figures~\ref{fig:HMM_2_Q} and~\ref{fig:HMM_2_W} provide an overview of the 
behaviour of the Q-2 and Q-2-M HMM over the period 2003-2018. We chose those two aggregation levels as they best illustrate the interest of using HMMs to detect patterns and the effect of temporal resolution change. The top panel of each figure reports three elements: (i) the realised operational losses 
(grey bars), (ii) the empirical 90th percentile of historical losses (dashed line), and 
(iii) the one-step-ahead 90th quantile prediction produced by the two-state HMM and by the HMM including the VSTOXX index as a covariate. The middle panels show the filtered state probabilities (for $s_{t}=1$) : the interpretation is natural, the state $s_{t}=0$ represents a low-loss regime whereas the other a high-loss regime. These probabilities indicate how frequently the model switches between these two regimes, and whether the inclusion of the covariate leads to more (or less) persistent 
state durations.
The bottom panel displays the VSTOXX index, used here as a proxy for financial stress. We would like to know whether the inclusion of VSTOXX improves regime detection. This series helps assess whether the inferred high-loss periods coincide with episodes of elevated market volatility, especially when the covariate is explicitly incorporated into 
the transition structure. \newline

\FloatBarrier

\FloatBarrier

\noindent First, both figures show that operational losses related to internal frauds exhibit regime dependence. Periods of relatively low and stable losses alternate with periods characterized by higher and more clustered losses. For instance, the 2007–2009 financial crisis can be identified as a cluster of high losses. Such behavior is not aligned with the assumptions underlying the traditional LDA framework, in which losses are modeled as a compound Poisson process with constant intensity. Under this framework, risk metrics such as the $90^{\text{th}}$ percentile considered here are time-invariant. This is illustrated by the historical $90^{\text{th}}$ percentile shown in the figures (dashed line), which provides a benchmark across the sample. However, this value is systematically misaligned with periods of high or low losses, highlighting a well-known limitation of stationary LDA models: they tend to overestimate risk during low-loss periods and underestimate it during crisis periods, mainly because they ignore regime shifts. Similar conclusions are drawn in \citet{dionne2016hidden}, who show the presence of distinct loss regimes in bank operational loss data. \newline

\noindent Although the specification of the model is identical in both cases, the figures show that the standard two-state HMM is able to capture the regime structure at low frequency but struggles to do so at higher frequency. As the statistical properties of the loss series differ substantially across levels of time aggregation, this leads to different regime dynamics. At the quarterly level, the filtered probabilities remain in state 0 or 1 over several consecutive quarters, indicating strong persistence. The resulting regimes are economically interpretable and align with major financial stress periods. By contrast, at the weekly level, the same model displays a different pattern, rapidly alternating between states, with filtered probabilities switching frequently from one week to the next and implying very short regime durations. This behavior suggests that, at higher frequency, the homogeneous HMM reacts primarily to short-term fluctuations rather than to long-term structural changes \cite{cappe2005inference}. This should not be interpreted as an absence of regimes at the weekly level; rather, it highlights a limitation of the homogeneous HMM when estimated on noisy, high-frequency data. In the absence of additional information, the model uses frequent state changes to accommodate high variability in the observations and fails to capture the overall regime structure. This phenomenon is well documented in the literature. \newline

\noindent Finally, introducing the VSTOXX index changes the dynamics of the inferred states, particularly at the weekly frequency. The filtered probabilities become more persistent, with less alternation between states, indicating that the model is now able to detect regime persistence. At the quarterly frequency, the impact of including the macroeconomic covariate is less pronounced, as regime persistence is already well captured by the standard HMM. The macroeconomic variable should not be interpreted as a direct driver of operational losses; its role is purely conditional and serves to provide additional information that improves the identification of regime transitions. \newline

\noindent Overall, both figures support the view that operational risk losses are generated by a regime-switching process that can be sensitive to the macro-financial environment in the case of internal fraud losses. Furthermore, these results highlight the relevance of incorporating an additional variable when modeling noisy, high-frequency observations. \newline

\noindent Beyond the visual inspection of regime dynamics, a more systematic assessment of model performance is provided in Table~\ref{tab:mse_internal_fraud}. The table reports the Mean Squared Error (MSE) between the predicted and observed $90^{\text{th}}$ quantile computed on exceedances for models with two, three, and four hidden states, estimated at quarterly, monthly, and weekly frequencies, with or without the inclusion of the macroeconomic covariate. \newline

\begin{table}[htbp]
\centering
\footnotesize
\renewcommand{\arraystretch}{1.1}
\setlength{\tabcolsep}{4pt}
\begin{tabular}{lcccccc}
\toprule
 & \multicolumn{2}{c}{\textbf{Quarterly}} 
 & \multicolumn{2}{c}{\textbf{Monthly}} 
 & \multicolumn{2}{c}{\textbf{Weekly}} \\
\cmidrule(lr){2-3} \cmidrule(lr){4-5} \cmidrule(lr){6-7}
\textbf{Model} 
 & \makecell{\textbf{Excl.}\\\textbf{VSTOXX}} 
 & \makecell{\textbf{Incl.}\\\textbf{VSTOXX}}
 & \makecell{\textbf{Excl.}\\\textbf{VSTOXX}} 
 & \makecell{\textbf{Incl.}\\\textbf{VSTOXX}}
 & \makecell{\textbf{Excl.}\\\textbf{VSTOXX}} 
 & \makecell{\textbf{Incl.}\\\textbf{VSTOXX}} \\
\midrule
2-State & 4.27 & 3.77 & 2.86 & 1.84 & 1.55  & 0.923 \\
3-State & 6.77 & 3.94 & 2.88 & 2.01 & 1.13  & 0.887 \\
4-State & 2.59 & 3.06 & 2.43 & 2.26 & 0.982 & 0.859 \\
\bottomrule
\end{tabular}
\caption{Mean Squared Error (MSE) between predicted and observed $90^{\text{th}}$ quantile exceedances for Internal Fraud (ET1), by aggregation level, number of hidden states, and inclusion of the macro-financial variable (VSTOXX).}
\label{tab:mse_internal_fraud}
\end{table}

\noindent A first observation concerns the number of states used in the model. At quarterly or monthly frequencies, where the number of available observations is limited, increasing the number of hidden states leads to instability. Two- and three-state models are able to identify the main regimes, whereas the four-state specification exhibits excessive regime switching and overfits the data. This reflects the bias–variance trade-off in hidden Markov models: while additional states increase flexibility, they may also lead to overfitting and reduced interpretability. Time aggregation also plays a key role in regime inference. Quarterly aggregation naturally smooths short-term fluctuations, facilitating the identification of regimes using a homogeneous HMM. At finer aggregation levels, the same model focuses on short-term variability and is unable to detect the main regime switches. Finally, the inclusion of the macroeconomic covariate improves the predictions provided by the standard HMM in all two- and three-state configurations. By providing additional structure, the macroeconomic covariate helps distinguish between short-term fluctuations in losses and genuine regime shifts, and can partially substitute for an increase in the number of hidden states by stabilizing regime inference at high frequencies. \newline

\subsection{Experimental results on the other categories}

Table~\ref{mse_all_risk} reports the Mean Squared Error (MSE) computed on exceedances of the $90^{\text{th}}$ quantile for all event types, across different time aggregations, numbers of hidden states, and with or without macroeconomic covariate. \newline

\noindent For event types ET3 (Employment Practices) and ET4 (Clients, Products and Business Practices), the loss series do not exhibit visible regime changes. As a result, the application of HMMs does not lead to a meaningful improvement in predictive performance for these categories. For the remaining event types, the results reported in Table~\ref{mse_all_risk} reveal heterogeneous behaviours across risk categories. In particular, for ET5 (Physical Damage) and ET6 (System Failures), the inclusion of the VSTOXX does not systematically improve prediction accuracy across aggregation levels or state-space dimensions. In several configurations, models without the macroeconomic covariate achieve comparable or even lower MSE values. This outcome is economically plausible, as losses related to physical damage or system failures are not directly expected to be linked to financial market volatility as captured by the VSTOXX index. These risks are more likely driven by operational, technological, or environmental factors. Nevertheless, the relevance of alternative covariates, such as climate-related indicators, could be explored in future work. By contrast, for internal fraud (ET1) and external fraud (ET2), the inclusion of the VSTOXX variable leads to lower MSE values in most specifications. This suggests that, for these event types, operational losses exhibit a measurable dependency on the macroeconomic environment, which can be exploited within a regime-switching framework. \newline


\begin{table}[!ht]
\centering
\scriptsize
\setlength{\tabcolsep}{4pt}
\renewcommand{\arraystretch}{1.1}
\begin{tabular}{lcccccc}
\toprule
& \multicolumn{2}{c}{\textbf{Quarterly}} 
& \multicolumn{2}{c}{\textbf{Monthly}} 
& \multicolumn{2}{c}{\textbf{Weekly}} \\
\cmidrule(lr){2-3} \cmidrule(lr){4-5} \cmidrule(lr){6-7}
& \makecell{\textbf{Excl.}\\\textbf{VSTOXX}} 
& \makecell{\textbf{Incl.}\\\textbf{VSTOXX}} 
& \makecell{\textbf{Excl.}\\\textbf{VSTOXX}} 
& \makecell{\textbf{Incl.}\\\textbf{VSTOXX}} 
& \makecell{\textbf{Excl.}\\\textbf{VSTOXX}} 
& \makecell{\textbf{Incl.}\\\textbf{VSTOXX}} \\
\midrule
\multicolumn{7}{c}{\textbf{ET1 -- Internal Fraud}} \\
\midrule
2-State & 4.27 & 3.77 & 2.86 & 1.84 & 1.55 & 0.923 \\
3-State & 6.77 & 3.94 & 2.88 & 2.01 & 1.13 & 0.887 \\
4-State & 2.59 & 3.06 & 2.43 & 2.26 & 0.982 & 0.859 \\
\midrule
\multicolumn{7}{c}{\textbf{ET2 -- External Fraud}} \\
\midrule
2-State & 450 & 460 & 67 & 64 & 20 & 24 \\
3-State & 564 & 570 & 79 & 52 & 22 & 23 \\
4-State & 208 & 531 & 73 & 65 & 22 & 24 \\
\midrule
\multicolumn{7}{c}{\textbf{ET5 -- Physical Damage}} \\
\midrule
2-State & 1.64 & 1.73 & 0.53 & 0.63 & 0.36 & 0.31 \\
3-State & 0.40 & 1.30 & 0.59 & 0.67 & 0.34 & 0.30 \\
4-State & 0.24 & 1.12 & 0.31 & 0.77 & 0.33 & 0.28 \\
\midrule
\multicolumn{7}{c}{\textbf{ET6 -- System Failures}} \\
\midrule
2-State & 19.1 & 19.5 & 3.56 & 4.18 & 0.73 & 0.71 \\
3-State & 13.6 & 18.3 & 2.52 & 3.48 & 0.67 & 0.71 \\
4-State & 10.7 & 17.0 & 3.06 & 3.24 & 0.68 & 0.72 \\
\midrule
\multicolumn{7}{c}{\textbf{ET7 -- Process Management}} \\
\midrule
2-State & -- & -- & 61.9 & 59.5 & -- & -- \\
3-State & -- & -- & 69.2 & 48.9 & -- & -- \\
4-State & -- & -- & 61.4 & 41.3 & -- & -- \\
\bottomrule
\end{tabular}
\caption{MSE values between the predicted and observed $90^{\text{th}}$ quantile exceedances across event types, aggregation levels, and HMM specifications. A dash indicates that the model could not be calibrated.}
\label{mse_all_risk}
\end{table}


\noindent Overall, this study shows that multivariate HMMs are able to capture economically meaningful dependencies between the macroeconomic environment and operational losses for certain event types, while remaining less informative for others. These results indicate that the sensitivity of operational risk to macroeconomic conditions is event-type specific and cannot be assumed a priori. From a stress-testing perspective, the proposed framework provides a flexible tool to explore dependencies between operational losses and a wide range of covariates, including macroeconomic, financial, or climate-related variables. \newline

\FloatBarrier
\newpage
\section{Conclusion}

\noindent The Hidden Markov Model (HMM) framework provides a dynamic approach to modeling operational risk, allowing for the inclusion of macroeconomic variables such as the VSTOXX. A key finding of this study is that macroeconomic information can be incorporated in a statistically meaningful way for some operational risk event types. In particular, when the loss series exhibits regime-dependent behavior that is aligned with the macro-financial environment, the multivariate HMM specification (including the VSTOXX) provides more accurate predictions than the standard HMM. Conversely, for event types for which no such regime structure is visible, the inclusion of the macroeconomic covariate does not systematically improve performance. These results suggest that some event types are more sensitive to macroeconomic variables than others. While the Gaussian assumption simplifies computations, it may not be sufficient for capturing extreme operational losses. An extension to heavy-tailed distributions, such as the Student-$t$ or other elliptical families \citep{McNeilFreyEmbrechts2015}, would therefore provide a more flexible representation of tail risk. Additionally, further analysis could explore the inclusion of alternative macroeconomic indicators, such as interest rates, to determine whether other financial variables offer better predictions for specific categories of operational risk. Overall, this study represents a first step to analyze the dependence between macroeconomic variables and operational risk losses, and provides a flexible framework to assess when and for which event types the inclusion of macroeconomic covariates is empirically justified. Note that the framework can be extended to include other covariates that are increasingly considered in regulatory stress-testing exercises, such as climate-related or geopolitical and cyber risk indicators.

\bibliographystyle{chicago}
\bibliography{mybibliography}

@misc{basel2006ii,
  author      = {{Basel Committee on Banking Supervision}},
  title       = {International Convergence of Capital Measurement and Capital Standards: A Revised Framework --- Comprehensive Version},
  institution = {Bank for International Settlements},
  year        = {2006},
  url         = {https://www.bis.org/publ/bcbs128.pdf}
}

@misc{basel2011iii,
  author      = {{Basel Committee on Banking Supervision}},
  title       = {{Basel III}: A Global Regulatory Framework for More Resilient Banks and Banking Systems},
  institution = {Bank for International Settlements},
  year        = {2011},
  url         = {https://www.bis.org/publ/bcbs189.pdf}
}

@misc{basel2017iv,
  author      = {{Basel Committee on Banking Supervision}},
  title       = {Finalising {Basel III}: In Brief},
  institution = {Bank for International Settlements},
  year        = {2017},
  url         = {https://www.bis.org/bcbs/publ/d424_inbrief.pdf}
}

@article{lindskog2003,
  author    = {Lindskog, Filip and McNeil, Alexander J.},
  title     = {Common Poisson Shock Models: Applications to Insurance and Credit Risk},
  journal   = {ASTIN Bulletin: The Journal of the IAA},
  volume    = {33},
  number    = {2},
  pages     = {209--238},
  year      = {2003},
  publisher = {Cambridge University Press}
}

@book{McNeilFreyEmbrechts2015,
  author    = {McNeil, Alexander J. and Frey, R{"u}diger and Embrechts, Paul},
  title     = {Quantitative Risk Management: Concepts, Techniques and Tools},
  publisher = {Princeton University Press},
  address   = {Princeton, NJ},
  edition   = {Revised},
  year      = {2015}
}

@article{kreinin2017correlated,
  title={Correlated Poisson Processes},
  author={Kreinin, Alexander},
  year={2017},
  month={December},
}

@article{chiu2017correlated,
  title={Correlated Multivariate Poisson Processes and Extreme Measures},
  author={Chiu, Michael and Jackson, Kenneth and Kreinin, Alexander},
  journal={Model Assisted Statistics and Applications},
  volume={12},
  year={2017},
  month={February}
}

@article{bae2017backward,
  title={A Backward Construction and Simulation of Correlated Poisson Processes},
  author={Bae, Taehan and Kreinin, Alexander},
  journal={Journal of Statistical Computation and Simulation},
  volume={87},
  pages={1--15},
  year={2017},
  month={January}
}

@article{duch2014new,
  title={New Approaches to Operational Risk Modeling},
  author={Duch, Konrad and Jiang, Yijun and Kreinin, Alexander},
  journal={IBM Journal of Research and Development},
  volume={58},
  number={3},
  pages={3:1--3:9},
  year={2014},
  month={July},
  publisher={IBM}
}

@article{tankov2016levy,
  title={L{\'e}vy Copulas: Review of Recent Results},
  author={Tankov, Peter},
  year={2016},
  month={December},
}

@article{tankov2019simulation,
  title={Simulation and Option Pricing in L{\'e}vy Copula Models},
  author={Tankov, Peter},
  year={2019},
  month={October},
}

@article{boecker2010multivariate,
  title={Multivariate Models for Operational Risk},
  author={B{\"o}cker, Klaus and Kl{\"u}ppelberg, Claudia},
  journal={Quantitative Finance},
  volume={10},
  pages={855--869},
  year={2010},
  month={October},
  publisher={Taylor \& Francis}
}

@article{dionne2016hidden,
  title={Hidden Markov Regimes in Operational Loss Data: Application to the 2007–2009 Financial Crisis},
  author={Dionne, Georges and Saissi Hassani, Saeed},
  journal={CIRRELT},
  year={2016}
}

@article{fung2019multivariate,
  title={Multivariate Cox Hidden Markov Models with an Application to Operational Risk},
  author={Fung, Timothy C and Badescu, Alexandru L and Lin, X Sheldon},
  journal={Scandinavian Actuarial Journal},
  volume={2019},
  number={2},
  pages={93--118},
  year={2019},
  publisher={Taylor \& Francis}
}

@article{hambuckers2018markov,
  title={A Markov-Switching Generalized Additive Model for Compound Poisson Processes, with Applications to Operational Loss Models},
  author={Hambuckers, Julien and Kneib, Thomas},
  journal={Journal of Business \& Economic Statistics},
  volume={36},
  number={4},
  pages={610--622},
  year={2018},
  publisher={Taylor \& Francis}
}

@book{chernobai2011operational,
  title = {Operational Risk: A Guide to Basel II Capital Requirements, Models, and Analysis},
  author = {Chernobai, Anna S. and Rachev, Svetlozar T. and Fabozzi, Frank J.},
  year = {2011},
  publisher = {Wiley},
  address = {Hoboken, NJ},
}

@book{CrouhyGalaiMark2014,
  title = {The essentials of Risk Management, second edition},
  author = {Michel Crouhy and Dan Galai and Robert Mark},
  year = {2014},
  publisher = {McGraw Hill},
  address = {New York},
}

@article{cope2012macroenvironmental,
  title = {Macroenvironmental Determinants of Operational Losses},
  author = {Cope, Eric W. and Piche, Michael T. and Walter, John S.},
  journal = {Journal of Banking \& Finance},
  volume = {36},
  number = {5},
  pages = {1362--1380},
  year = {2012},
}

@article{hess2011impact,
  title = {The Impact of the Financial Crisis on Operational Risk in Banks},
  author = {Hess, Christoph},
  journal = {Journal of Operational Risk},
  volume = {6},
  number = {1},
  pages = {23--35},
  year = {2011},
  doi = {10.21314/JOP.2011.098},
}

@article{fiordelisi2014determinants,
  title = {The Determinants of Reputational Risk in the Banking Sector},
  author = {Fiordelisi, Franco and Soana, Maria Grazia and Schwizer, Paola},
  journal = {Journal of Banking \& Finance},
  volume = {37},
  number = {5},
  pages = {1359--1371},
  year = {2014},
}

@article{ChavezDemoulin2016,
  author = {Chavez-Demoulin, Valérie and Embrechts, Paul and Hofert, Marius},
  title = {An Extreme Value Approach for Modeling Operational Risk Losses Depending on Covariates},
  journal = {Journal of Risk and Insurance},
  volume = {83},
  number = {3},
  pages = {735-776},
  year = {2016},
  doi = {10.1111/jori.12072},
}

@Article{EBA21,
author = {{European Banking Authority}},
title = {2021 EU-wide stress test – methodological note},
note = {Retrieved on July 19, 2021 on \url{ https://www.eba.europa.eu/eba-launches-2021-eu-wide-stress-test-exercise}},
year = {2021},
month = {January}
}

@Article{STAMPE2017,
author = {{European Central Bank}},
title = {Stress-Test Analytics for Macroprudential Purposes in the euro area},
note = {Retrieved on April 26, 2025 on \url{https://www.ecb.europa.eu/press/conferences/shared/pdf/20170511_2nd_mp_policy/DeesHenryMartin-Stampe-Stress-Test_Analytics_for_Macroprudential_Purposes_in_the_euro_area.en.pdf}},
year = {2017},
month = {February}
}

@Article{BCBS05,
author = {{Basel Committee on Banking Supervision}},
title = "An Explanatory Note on the {Basel II IRB} Risk Weight Functions",
note = {Retrieved on July 19, 2021 on \url{https://www.bis.org/bcbs/irbriskweight.htm}},
year = "2005"
}

@article{rabiner1989,
  author    = {L. R. Rabiner},
  title     = {A Tutorial on Hidden Markov Models and Selected Applications in Speech Recognition},
  journal   = {Proceedings of the IEEE},
  year      = {1989},
  volume    = {77},
  number    = {2},
  pages     = {257--286},
  doi       = {10.1109/5.18626}
}

@book{bishop2006pattern,
  author    = {Christopher M. Bishop},
  title     = {Pattern Recognition and Machine Learning},
  publisher = {Springer},
  year      = {2006},
  address   = {New York},
  pages     = {456, 486}
}

@book{Jorion2011,
  author    = {Philippe Jorion},
  title     = {Financial Risk Manager Handbook},
  publisher = {Wiley},
  year      = {2011},
  address   = {New Jersey}
}

@book{CruzPetersShevchenko2015,
author = {Marcelo G. Cruz and Gareth W. Peters and Pavel V. Shevchenko},
publisher = {John Wiley \& Sons, Ltd},
isbn = {9781118573013},
title = {Fundamental Aspects of Operational Risk and Insurance Analytics: A Handbook of Operational Risk},
doi = {https://doi.org/10.1002/9781118573013.biblio},
year = {2015}
}

@book{murphy2012machine,
  title     = {Machine Learning: A Probabilistic Perspective},
  author    = {Kevin P. Murphy},
  year      = {2012},
  publisher = {MIT Press},
  address   = {Cambridge, MA},
  pages     = {431--432}
}

@article{bastide2023derivatives,
  author    = {Dorinel Bastide and Stéphane Crépey and Samuel Drapeau and Mekonnen Tadese},
  title     = {Derivatives Risks as Costs in a One‑Period Network Model},
  journal   = {Frontiers of Mathematical Finance},
  year      = {2023},
  volume    = {2},
  number    = {3},
  pages     = {283--312},
  doi       = {10.3934/fmf.2023014}
}

@book{bellini2016stress,
  author    = {Tiziano Bellini},
  title     = {Stress Testing and Risk Integration in Banks: A Statistical Framework and Practical Software Guide},
  publisher = {Elsevier},
  year      = {2016},
  edition   = {1},
  isbn      = {978-0128035900}
}

@book{cappe2005inference,
  author    = {Olivier Capp{\'e} and Eric Moulines and Tobias Ryden},
  title     = {Inference in Hidden Markov Models},
  publisher = {Springer},
  year      = {2005},
  series    = {Springer Series in Statistics},
  address   = {New York},
  doi       = {10.1007/0-387-28982-8}
}

@book{mclachlan2000finite,
  author    = {Geoffrey J. McLachlan and David Peel},
  title     = {Finite Mixture Models},
  publisher = {Wiley},
  year      = {2000},
  series    = {Wiley Series in Probability and Statistics},
  address   = {New York},
  isbn      = {978-0471006268}
}

@book{tukey1977eda,
  author    = {John W. Tukey},
  title     = {Exploratory Data Analysis},
  publisher = {Addison-Wesley},
  year      = {1977},
  address   = {Reading, MA},
  isbn      = {978-0201076165}
}

@article{zhang2001segmentation,
  author    = {Zhang, Y. and Brady, M. and Smith, S.},
  title     = {Segmentation of Brain MR images through a hidden Markov random field model and the expectation–maximization algorithm},
  journal   = {IEEE Transactions on Medical Imaging},
  year      = {2001},
  volume    = {20},
  number    = {1},
  pages     = {45--57}
}

\end{document}